\documentclass[prl,twocolumn,showpacs,preprintnumbers,amsmath,amssymb]{revtex4}

\usepackage{graphicx}
\usepackage{dcolumn}
\usepackage{bm}

\begin{document}

\preprint{APS/123-QED}

\title{Mapping of spin lifetimes to electronic states in \emph{n}-type GaAs near the metal-insulator transition}

\author{L. Schreiber}
\author{M. Heidkamp}
\author{T. Rohleder}
\author{B. Beschoten}
\email{beschoten@physik.rwth-aachen.de}
\author{G. G\"untherodt}
\affiliation{{II.} Physikalisches Institut, and Virtual Institute
for Spin Electronics (ViSel), RWTH Aachen University, Templergraben
55, 52056 Aachen, Germany}

\date{\today}

\begin{abstract}

The longest spin lifetimes in bulk \emph{n}-GaAs exceed 100 ns for
doping concentrations near the metal-insulator transition (J.M.
Kikkawa, D.D. Awschalom, Phys. Rev. Lett. 80, 4313 (1998)). The
respective electronic states have yet not been identified. We
therefore investigate the energy dependence of spin lifetimes in
\emph{n}-GaAs by time-resolved Kerr rotation. Spin lifetimes vary
by three orders of magnitude as a function of energy when
occupying donor and conduction band states. The longest spin
lifetimes ($>100$ ns) are assigned to delocalized donor band
states, while conduction band states exhibit shorter spin
lifetimes. The occupation of localized donor band states is
identified by short spin lifetimes ($\sim300$ ps) and a distinct
Overhauser shift due to dynamic nuclear polarization.

\end{abstract}

\maketitle

Within the framework of the emerging field of spintronics, the
spin degree of freedom is exploited for information storage as
well as processing and could serve as a qubit for quantum
computation \cite{Awschalom07}. Spin coherence and long spin
lifetimes are a prerequisite for novel spintronic devices.
Electron spins in Si-doped bulk \emph{n}-GaAs drew attention, when
long spin lifetimes $T_2^*> 100$~ns and coherence lengths larger
than 100~$\mu$m were determined using time-resolved Faraday
rotation \cite{Kikkawa98,Kikkawa99}. Since then \emph{n}-GaAs was
used as a model system to investigate spin injection and spin
transport phenomena \cite{Kato04,Crooker05,Lou07}. The long
$T_2^*$ of bulk \emph{n}-GaAs, however, is restricted to a doping
concentration in the vicinity of the metal-insulator transition
(MIT) and shortens dramatically towards both sides of the
transition \cite{Kikkawa98, Dzhioev02}. Similar results in
\emph{n}-type GaN \cite{Beschoten01}, and \emph{n}-type Si
\cite{Zarifis87} point to a universal phenomenon. However, the
respective electronic states yielding these long $T_2^*$ near the
MIT have not been identified so far.

Various spin relaxation mechanisms have been considered to explain
the dependence of $T_2^*$ on carrier concentration, temperature,
and magnetic field $B$. The relevant relaxation mechanisms differ
substantially for delocalized spins with, e.g., the
D'yakonov-Perel' (DP) dephasing mechanism
\cite{Fabian99,Song02,Yu05,Shklovskii06}, and for spins localized
at impurity sites with, e.g., relaxation by hyperfine interaction
\cite{Shklovskii06,Dzhioev01,Putikka04}. Concerning the electronic
states of \emph{n}-type semiconductors, the MIT was shown to occur
within the donor band (DB) \cite{Romero90}, which is separated
from the conduction band (CB). Near the Fermi level ($E_F$), the
electronic structure is governed by both doping induced disorder
and local Coulomb correlation. The former yields
Anderson-localized states in the upper and lower donor band-tails,
which are separated from extended states in the center by mobility
edges \cite{Anderson58}. The latter may lead to a Coulomb gap $U$
at $E_F$ \cite{Efros84}. Both interactions yield a complex
electronic structure with coexisting localized and delocalized DB
states as well as CB states. Spin dephasing in \emph{n}-GaAs has
mostly been investigated for states at $E_F$
\cite{Kikkawa98,Kikkawa99}. There is, however, no energy-resolved
study of $T_2^*$, which would allow to assign spin lifetimes to
the respective electronic states of both the donor and the
conduction band. We expect that this assignment helps to identify
the dominant spin relaxation mechanisms in the vicinity of the
MIT.

In this Letter, we study the spin lifetime $T_2^*$ of coherent
electron spin states in \emph{n}-GaAs, which are optically excited
in both the donor and conduction band and probed by time-resolved
Kerr rotation (TRKR) at 6~K. Due to the coexistence of distinct
electronic states, the sample is not characterized by a single
$T_2^*$: $T_2^*$ varies by three orders of magnitude as a function
of photon energy. The longest $T_2^*$ values which may exceed 100~ns
are found for delocalized donor band states, while free conduction
band states exhibit shorter spin lifetimes. Our time-resolved Kerr
signal shows up to three exponential decay regimes with different
precession frequencies. The latter can change due to an additional
nuclear magnetic field arising from dynamically polarized nuclei,
when resonantly pumping spins into localized DB states.

Two (001)-oriented, 500~$\mu$m thick GaAs wafers with different Si-doping
concentrations have been investigated: Sample A with a carrier
concentration of $(2.4 \pm 0.2) \times 10^{16}$~cm$^{-3}$ is doped close to
the MIT (critical carrier concentration in Si:GaAs $n_c \cong 1.5 \times
10^{16}$~cm$^{-3}$) \cite{Romero90}. Reference sample B has a carrier
concentration of $(1.5 \pm 0.4) \times 10^{18}$~cm$^{-3}$ and is therefore
degenerated \cite{Efros84}. We used two tuneable, mode-locked
Ti:Al$_2$O$_3$ lasers providing $\sim$150~fs optical pulses corresponding
to a spectral width of $\approx 6$~nm at a repetition frequency of 80~MHz.
Electronic phase-locking of both lasers enables us to employ one laser for
spin pumping at an energy $E_{pu}$ and the other one for probing the spin
orientation at an energy $E_{pr}$ after a variable delay time $\Delta t =
0...16$~ns. The normal-incident pump pulses, which were circularly
polarized by a photo-elastic modulator (PEM), excite spin-polarized
electrons and holes oriented along the beam direction in the strain-free
mounted samples with an average power $\langle P \rangle = 50$~W/cm$^2$.
The projection of the pump induced spin magnetization onto the surface
normal of the sample is determined with linearly polarized laser pulses by
Faraday rotation $\theta_F$ in transmission and by Kerr rotation $\theta_K$
in reflection. Transverse magnetic fields $B$ are applied in the plane of
the sample.

\begin{figure}
\includegraphics{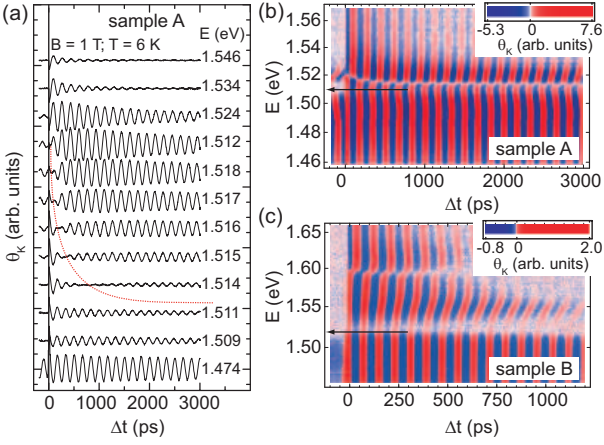}
\caption{\label{fig1} (Color) Time-resolved Kerr rotation for photon
energies $E=E_{pu}=E_{pr}$ at 6~K with (a) $\theta_K(\Delta t)$ for
sample A at $B=1$~T; the red line shows the shift of the beating
node; (b) $\theta_K(\Delta t,E)$ for sample A at $B=1$~T; (c)
$\theta_K(\Delta t,E)$ for the degenerate sample B at $B=6$~T.
Arrows mark the respective energies, above which the transmission
drops below 5\%.}
\end{figure}

In Figure \ref{fig1}(a), we plot $\theta_K(\Delta t)$ of sample A
measured for various photon energies $E=E_{pu}=E_{pr}$ at $B=1$~T
and at $T=6$~K. Obviously, the spins precess at all energies $E$,
but the damping of the oscillations and thus $T_2^*$ is
$E$-dependent: for $E$ below the CB edge $T_2^*$ is long, whereas
for the highest $E$, at which CB states are pumped, $T_2^*$ is much
reduced. Strikingly, a node in the oscillation envelope (red line in
Fig. \ref{fig1}(a)) near the band edge indicates that the spins
precess with at least two Larmor frequencies $\omega^{(i)}$.
Therefore, $\theta_K(\Delta t)$ is described by $n$ decay components

\begin{equation}\label{fitformula}
  \theta_K(\Delta t) = \sum_i^n A^{(i)} \exp\left(-\frac{\Delta t}{T_2^{*(i)}}\right)\cos\left(\omega^{(i)} \Delta t +
  \phi^{(i)}\right),
\end{equation}

\noindent where $A^{(i)}$, $T_2^{*(i)}$ and $\phi^{(i)}$ are the
initial amplitude, the spin lifetime and the phase of component
$i$, respectively. Since this beating node shifts with increasing
$E$ towards $\Delta t=0$, the difference of the two $\omega^{(i)}$
increases. The Larmor frequencies $\omega^{(i)}=g^{*(i)} \mu_B B /
\hbar$ depend upon component-specific effective g-factors
$g^{*(i)}$. Note that the pump-induced spin packets, which exhibit
a \emph{continuous} distribution of g-factors, precess with only
one energetically averaged $g^*$ but with reduced $T_2^*$ due to
inhomogeneous dephasing \cite{Kikkawa98}. Thus, electronic states
with two g-factor dispersions like, e.g., DB and CB states are
pumped simultaneously to yield separated $\omega^{(i)}$. Figure
\ref{fig1}(b) displays $\theta_K(\Delta t,E)$ of sample A in a
false color plot. The $E$-dependence of $\omega^{(i)}$ is obvious
from horizontal variations of maxima and minima of $\theta_K$ and
most distinct at the band edge (horizontal arrow). Non-vanishing
$\theta_K$ at $\Delta t<0$ evidence long $T_2^*\gtrsim 6$~ns,
since the signal stems from spins exited by the previous pump
pulse. In Figure \ref{fig1}(c), this result is compared to
$\theta_K(\Delta t,E)$ of the degenerate sample B, for which $E_F$
lies well within the CB. For this sample, both $T_2^*$ and $g^*$
depend strongly on $E$ in the absorbing energy regime. However,
there are no nodes in the oscillation envelope at any energy (see
Fig. \ref{fig1}(c)). Note that the identity of DB states is
completely lost in this degenerate sample. This indicates that the
multi-component decay as well as the nodes in the oscillation
amplitude of sample A may arise from distinct DB and CB states.

\begin{figure}
\includegraphics{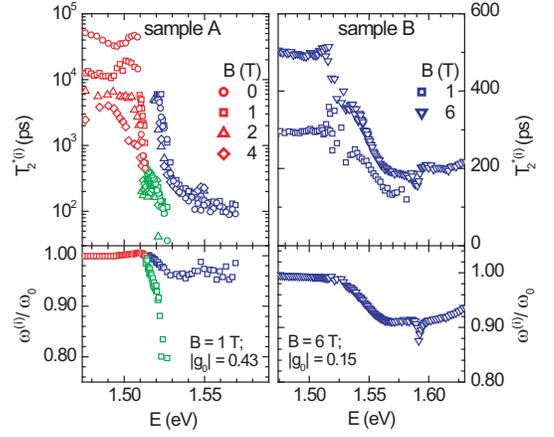}
\caption{\label{fig2} (Color) Fitted spin lifetimes $T_2^{*(i)}$ and
Larmor frequencies $\omega^{(i)}$ as a function of photon energy
$E=E_{pu}=E_{pr}$ for MIT sample A (left) and degenerate sample B
(right) at various magnetic fields $B$. The colors mark three energy
regions as discussed in the text. The Larmor frequencies are
normalized to $\omega_0$ calculated from $|g_0|=0.43$ and
$|g_0|=0.15$ for sample A and B, respectively.}
\end{figure}

To further support these findings, we plot in Figure \ref{fig2}
both $T_2^{*(i)}(E)$ and $\omega^{(i)}(E)$ values as determined
from least-squares fits of $\theta_K(\Delta t,E)$ according to Eq.
\ref{fitformula} for both samples and for various $B$ fields. The
values of long $T_2^*>20$~ns observed at $B=0$~T are extracted
from fits of resonant spin amplification scans \cite{Kikkawa98}.
For sample A, $T_2^*$ drops by more than 3 orders of magnitude as
a function of photon energy $E$, however, three energy regions can
be distinguished marked by red, green and blue symbols. At low $E$
(red), $T_2^*$ is nearly independent of $E$ but decreases rapidly
with the increase of $B$. In this regime we also observe $T_2^*
\approx 110$~ns at $B=0$~T at reduced pump power $\langle P
\rangle = 10$~W/cm$^2$ (not shown) similar to Ref.
\cite{Kikkawa98}. Approaching the CB edge $T_2^*$ drops at all
$B$, but the Larmor frequency for the whole range (red) is almost
constant and exhibits $|g^*|=|g_0|=0.43$. Near the CB edge
(green), $B$-independent values of $T_2^* \sim 100...300$~ps are
determined, which distinguish themselves by a rapid decrease of
$\omega(E)$. At the beginning of the third region (blue), $T_2^*$
sets-in at $\sim 6$~ns and decreases with increasing $E$. The
corresponding $\omega$ decreases slightly and saturates at high
$E$. The overlap in energy of the second and third region, which
occurs due to the spectral width of the laser pulses, generates
the node in the oscillation envelope shown in Figure \ref{fig1}
\cite{comment1}. For the degenerate sample B, $T_2^{*(i)}(E)$ can
be fitted by one component $i$. As expected
\cite{Kikkawa98,Dzhioev02}, $T_2^*$ is over-all shorter compared
to sample A and $T_2^*$ increases with the increase of $B$, which
is typical for the DP dephasing mechanism \cite{Fabian99}.

\begin{figure}
\includegraphics{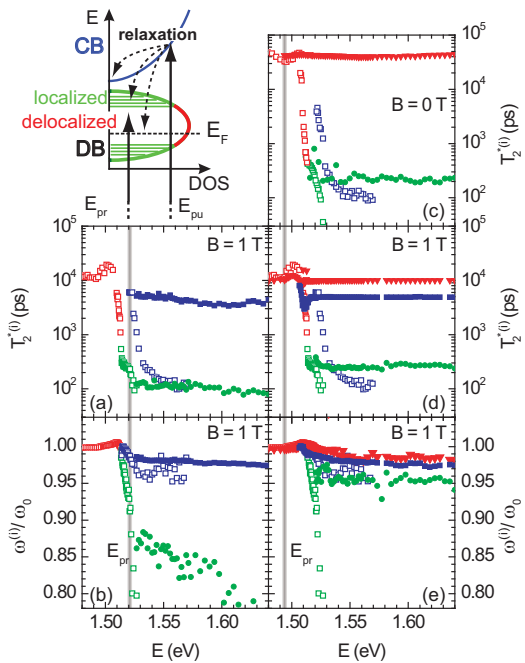}
\caption{\label{fig3} (Color) Upper left: sketch of density of DB
and CB states. Fitted spin lifetimes $T_2^{*(i)}$ and Larmor
frequencies $\omega^{(i)}$ as a function of pump laser energy
$E_{pu}$ at magnetic fields $B$ for MIT sample A (full symbols)
for different probe energies $E_{pr}$ is marked with vertical grey
lines. For comparison data from Fig. \ref{fig2} with
$E_{pu}=E_{pr}$ are also plotted (open symbols).}
\end{figure}

In the following, we assign the $T_2^{*(i)}$ of the spins observed in the
three energy ranges to carriers occupying different electronic states.
Since hole spins relax quickly $\lesssim 10$~ps in GaAs and excitons are
broken up at high magnetic fields $B \gtrsim 1$~T \cite{Oestreich96}, we
consider single electron states. Since the carrier concentration of sample
A is slightly above $n_c$, $E_F$ lies within the delocalized DB states as
sketched in Figure \ref{fig3}. Thus delocalized DB states are pumped at
lowest $E$, which exhibit the longest $T_2^*$ (red component in Fig.
\ref{fig2}). However, to our knowledge there is no relaxation mechanism of
spins, which accounts for their distinct $B$-dependence. We assign the
second energy range (green), which is missing for the degenerate sample B,
to Anderson localized electronic states at the DB tail. Their distinct
$E$-dependence of $\omega$ and $T_2^*$ can be linked to the decrease of
localization length upon approaching the DB tail. The localization of
electrons yields spin relaxation due to hyperfine interaction with the
nuclei \cite{Dzhioev01,Putikka04}. This gives rise to dynamic nuclear
polarization and a nuclear field, which alters $\omega$ (Overhauser shift)
\cite{Paget77}. However, from Ref. \cite{Dzhioev01} long $T_2^*\sim 300$~ns
are expected for localized electron spins, although we could not reproduce
this result with insulating \emph{n}-GaAs samples using TRKR. In Ref.
\cite{Putikka04}, an additional short $T_2^*\sim 100$~ps component is
predicted, when both localized and delocalized spins are pumped. This is
assigned to their cross-relaxation rate. Whereas this might explain our
short $T_2^*$, it does not account for the distinct $\omega(E)$ dependence.
In the third energy regime (blue), electrons are pumped in the CB. The
decrease of $\omega$ observed for both samples is due to $g^*$: The
absorption of the pump pulse lifts the local chemical potential, thus
reducing the absolute value of the energetically averaged $g^*$ factor
according to its dispersion in the CB \cite{Oestreich96}. The apparent
decrease of $T_2^*(E)$ might be explained by carrier cooling and interband
relaxation. Due to both, the carriers relax below the probed energy.

This notion can be confirmed by sweeping $E_{pu}$ with $E_{pr}$
held fixed at the bottom of the CB. The corresponding fits of
$T_2^{*(i)}(E_{pu})$ and $\omega^{(i)}(E_{pu})$ of sample A at
$B=1$~T are shown in Figure \ref{fig3}(a) and (b). Indeed, this
method allows to correct $T_2^*$ for energy relaxation, since
$T_2^*(E_{pu})$ (blue full symbols) decreases only slightly
compared to Figure \ref{fig2}. However, there is an additional
short $T_2^{*(i)}(E_{pu})$ (green full symbols), which results
from probing localized spins at the DB tail because of the
spectral width of the probe pulses. To clarify this point,
$E_{pr}$ is fixed at an energy even lower than the band gap (Fig.
\ref{fig3}(c)-(e)). For this $E_{pr}$, the longest
$T_2^{*(i)}(E_{pu})$ (red) attributed to delocalized DB states is
observed, which turned out to be nearly constant at $B = 0$~T.
Note that the average pump power is held constant, but the excited
carrier density changes by orders of magnitude in the absorbing
regime. This has a negligible effect on the longest $T_2^*$. More
strikingly, additional components $i$ become observable, when
$E_{pu}$ passes the localized and CB states. Thus, different
electronic states become occupied either due to direct optical
excitation or carrier relaxation as sketched in Figure
\ref{fig3}(upper left). From the onset of the $\theta_F$ signal
(not shown), we deduce that the delocalized DB states are occupied
within $\Delta t<10$~ps for all $E_{pu}$. At $B=1$~T (Fig.
\ref{fig3}(d)) and $E_{pu}$ beyond the band edge, the two long
$T_2^{*(i)}$ components (blue and red) generate nodes in the
oscillation envelope (Fig. \ref{fig1}(a)) and can be separated by
their $\omega^{(i)}$ (Fig. \ref{fig3}(e)). For fitting, however,
the longest $T_2^*$ (red) of delocalized DB had to be fixed with
minor influence on the blue component. The latter can be clearly
assigned to CB states by comparing it to $T_2^*(E_{pu})$ and
$g^*(E_{pu})$ of Figure \ref{fig3}(a) and (b). The existence of
three components $i$ and their onset when sweeping $E_{pu}$
confirms our assignment of $T_2^{*(i)}(E)$ to the three types of
electronic states.

\begin{figure}
\includegraphics{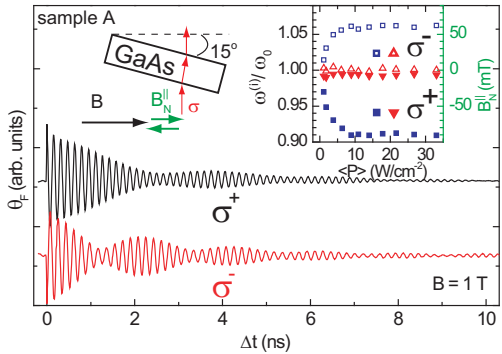}
\caption{\label{fig4} (Color) Faraday rotation $\theta_F(\Delta t)$ of
sample A for left ($\sigma^-$) and right ($\sigma^+$) circularly polarized
pump pulses (red arrows) at $B = 1$~T, $E_{pu}=1.514$~eV and
$E_{pr}=1.494$~eV. Inset: fitted normalized Larmor frequencies
$\omega^{(i)}$ and corresponding lateral nuclear magnetic fields $B_N^{\|}$
for both polarizations as a function of the average pump power.}
\end{figure}

Since an optically pumped spin imbalance of localized electrons
leads to pronounced dynamic nuclear polarization (DNP)
\cite{Paget77}, we finally check our assignment of electronic
states to $T_2^*$ by identifying DNP. When resonantly exciting
localized spins at, e.g., $E_{pu}=1.514$~eV, then DNP is optically
observed by the Overhauser shift. This shift results in a change
of $\omega$ due to the presence of a lateral nuclear magnetic
field $B_N^{\|}$ adding up to $B$. Since spins exhibiting long
$T_2^*$ are most sensitive to this shift, we chose the same
$E_{pr}$ as in Figure \ref{fig3}(c)-(e). In order to generate a
well-defined longitudinally pumped spin component and thus to
control the direction of $B_N^{\|}$ with respect to $B$, we
replaced the PEM by a quarter-waveplate and rotated sample A as
sketched in Figure \ref{fig4} \cite{Salis01}. In this geometry,
$\theta_F(\Delta t)$ exhibits nodes in the oscillation envelope at
long $\Delta t$, proving the presence of two long $T_2^{*(i)}$
components with different $\omega^{(i)}$. The dependence of
$\omega$ on the type of circularly polarization $\sigma^\pm$ of
the pump pulses, is clarified by the fitted $\omega^{(i)}$ in the
inset of Figure \ref{fig4}. The sign and magnitude of $B_N^{\|}$
is determined by $\sigma^\pm$ and the pump power, respectively.
$B_N^{\|}$ saturates for $\sigma^+$ at -90~mT, when $|g^*|=0.43$
is assumed to be constant. However, the blue component, which is
likely due to spins in the CB (cp. to Fig. \ref{fig3}(e)), is more
sensitive to $B_N$ than the spins attributed to delocalized DB
states (red). However, this point needs further investigation.
Since DNP is not observable when $E_{pu}$ is reduced below 1.5~eV,
our assignment of localized states is confirmed. The pronounced
$\omega^{(loc)}(E)$ dependence of the localized spins (green) (see
Fig. \ref{fig2} left) compared to the $\omega$ variation (blue) in
the inset of Figure \ref{fig4} suggests that the
$\omega^{(loc)}(E)$ is indeed influenced by increasing
localization and thus responsible for a rise of $B_N^{\|}$ at the
DB tail \cite{comment2}.

In summary, we have studied the energy dependence of spin
lifetimes in \emph{n}-type GaAs for electron doping near the
metal-insulator transition. Distinct spin lifetimes have been
assigned to both donor and conduction band states. Spin states at
the Fermi level are delocalized donor band states with the longest
spin lifetime, which may exceed 100~ns. The strong decrease of
spin lifetimes in the conduction band is related to energy
relaxation of hot electrons. Localized donor band states exhibit
the shortest spin lifetimes of $\sim$300~ps. Resonant optical
pumping of these localized states yields strong dynamic nuclear
polarization.

This work was supported by BMBF and by HGF.


\begin{thebibliography}{23}
\bibitem[{com({\natexlab{a}})}]{comment1}
\bibinfo{note}{The third region sets in at even lower energy $E\approx
  1.514$~eV, where $T_2^*$ cannot be determined exactly due to the beating.}
\bibitem[{com({\natexlab{b}})}]{comment2}
\bibinfo{note}{Even with the PEM, $\omega$ changes whithin $\sim 1$~min after switching
  on the pump. We attribute this to an Overhauser shift}.
\end{thebibliography}
\end{document}